\documentstyle[11pt,epsfig,epsf]{article}
\renewcommand{\theequation}{\arabic{equation}}
\newcommand{\bee}{\begin{equation}}
\newcommand{\ene}{\end{equation}}
\newcommand{\bea}{\begin{array}}
\newcommand{\ena}{\end{array}}
\newcommand{\beqa}{\begin{eqnarray}}
\newcommand{\enqa}{\end{eqnarray}}
\newcommand{\bean}{\begin{eqnarray*}}
\newcommand{\eean}{\end{eqnarray*}}

\def\ru1{\rule[-0.4truecm]{0mm}{1truecm}}

\def\up#1{\leavevmode \raise.16\hbox{#1}}
\def\sqr#1#2{{\vcenter{\vbox{\hrule height.#2pt
    \hbox{\vrule width.#2pt height#1pt \kern#1pt
      \vrule width.#2pt}
    \hrule height.#2pt}}}}

\newcounter{appendix}

\def\thebibliography#1{{\bf REFERENCES\markboth
 {REFERENCES}{REFERENCES}}\list
 {[\arabic{enumi}]}{\settowidth\labelwidth{[#1]}\leftmargin\labelwidth
 \advance\leftmargin\labelsep
 \usecounter{enumi}}
 \def\newblock{\hskip .11em plus .33em minus -.07em}
 \sloppy
 \sfcode`\.=1000\relax}

\begin{document}

\title{\hfill $%
\mbox{\small{
\begin{tabular}{r}
${\textstyle Bari-TH/98-308}$\\
${\textstyle hep-ph/9807572}$
\end{tabular}}} $ \\
[1truecm] A SEMIANALYTICAL METHOD TO EVOLVE PARTON DISTRIBUTIONS}
\author{Pietro SANTORELLI$^{a}$ and Egidio SCRIMIERI$^{b}$}
\date{$~$}

\maketitle

\vspace{-2.0truecm}

\begin{center}
{\it $^{a}$Dipartimento di Scienze Fisiche, Universit\`a "Federico
II" di Napoli,\\[0pt] and INFN, Sezione di Napoli\\[0pt] Mostra
d'Oltremare, Pad. 20, I-80125 Napoli, Italy.\\[0pt]
\vspace{.3truecm} $^{b}$Dipartimento di Fisica, Universit\`a di
Bari,\\[0pt] and INFN, Sezione di Bari\\[0pt] Via G. Amendola,
173, I-70126 Bari, Italy. }
\end{center}

\begin{abstract}
\noindent We present a new method to solve in a semianalytical way
the Dokshitzer-Gribov-Lipatov-Altarelli-Parisi evolution equations
at NLO order in the $x$-space. The method allows to construct an
evolution operator expressed in form of a rapidly convergent
series of matrices, depending only on the splitting functions.
This operator, acting on a generic initial distribution, provides
a very accurate solution in a short computer time (only a few
hundredth of second). As an example, we apply the method, useful
to solve a wide class of systems of integrodifferential equations,
to the polarized parton distributions.
\end{abstract}

\thispagestyle{empty}

\vspace{1truecm}


%\vspace{2truecm}
%\noindent
%
%\begin{tabular}{ll}
%{\bf Corresponding author}
%  & Pietro Santorelli, \\
%  & Dipartimento di Scienze Fisiche, Universit\`a di Napoli, \\
%  & Mostra d'Oltremare, Pad.20,\\
%  & I-80125, Napoli, Italy. \\
%E-mail: & Pietro.Santorelli@na.infn.it, Pietro.Santorelli@ba.infn.it\\
%Tel:    & +39 81 7253234 \\
%Fax:    & +39 81 2394508
%\end{tabular}

\newpage

\section{Introduction}

\label{s:intro}

The scaling violation of nucleon structure functions is described
in terms of Dokshitzer-Gribov-Lipatov-Altarelli-Parisi (DGLAP)
evolution equations \cite{DGLAP}. The DGLAP integrodifferential
equations describe the $Q^2$ dependence of the structure
functions, which are related, {\it via} the operator product
expansion, to the parton distributions for which the DGLAP
equations are usually written down. In this framework, the
analysis of the experimental data, is performed fixing at some
$Q_0^2$ the structure functions by assuming the parton
distributions and computing the convolutions with the coefficient
functions, which can be evaluated in perturbation theory. The
comparison with experimental data, which are distributed at
different values of $Q^2$, goes through the solution of DGLAP
equations for the parton distributions; thus a reliable and fast
algorithm to solve these equations is welcome.

In literature there are essentially three different approaches to
solve the DGLAP equations. The first one is based on the Laguerre
polynomials expansion \cite{FurmPetr}. This technique is quite
accurate up to $x$-values not smaller than $\bar{x}\approx
10^{-3}$; on the contrary, below $\bar{x}$ the convergence of the
expansion slows down \cite{kumLon,CorianoSavkli}. Given that
experimental data are already available down to about $\bar{x}$,
for the polarized case, and down to $10^{-5}$, for the unpolarized
case, this method results no longer practical (for a more detailed
discussion see section {\bf \ref{s:nunan}}).\newline An
alternative approach takes advantage of the fact that the moments
of the convolutions appearing in the equations factorize in such a
way that the analytical solution, in the momentum space, can be
written down \cite {Glucketal}. However, also in the most
favorable case in which the analytical expressions of the moments
of the initial conditions are known, the numerical Mellin
inversion is relatively CPU time consuming (see \cite
{SMCcomparison}).\newline Another strategy, the so called ``brute
force'' method \cite{bruteforce}, is fundamentally a
finite-differences method of solution. It reaches a good precision
in the small $x$-region \cite{Kumano,Fasching} but requires a
rather large amount of computer running time.

In this paper we present a new semianalytical method, to solve
DGLAP equations. It consists in constructing an evolution operator
which, depending only on the splitting functions, can be worked
out once for all. In this respect our strategy is similar to the
one in \cite {FurmPetr,CorianoSavkli}. Our method to perform the
convolutions instead, takes advantage of an $x$ discretization
(comparable to the one in \cite {Kumano,Fasching}) which allows us
to represent the evolution operator as a matrix. Thus the
procedure to construct the solution reduces merely to a
multiplication between the {\it evolution matrix} and an {\it initial vector}%
, and can be done in an extremely short computer time with the
required accuracy. This is particularly appealing in the analysis
of the experimental data on nucleon structure functions which
requires a large number of parton evolutions.\newline In the next
section we discuss the (formal) analytical solution of the DGLAP
equations; in the third one the algorithm to perform the
$x-$integration is presented. The last two sections are devoted to
analyze the numerical results relative to the evolution of
polarized parton distributions, to study the yield of our method
in comparison with others and to conclude.

\section{The Evolution Operator}

\label{s:tevolution}

In this section we review briefly the mathematical problem of DGLAP
evolution equations and propose a method to solve them. In the following we
limit ourselves to discuss the polarized parton distributions. The
application of our method to the unpolarized case is straightforward.

The DGLAP equation, up to Next-to-Leading-Order (NLO) corrections, for the
Non-Singlet distribution is
\begin{equation}
\frac{\partial}{\partial t} \Delta \tilde q_{NS}(x,t) = \left( \Delta \tilde
P^{(0)}_{NS}(x) + \alpha(t)\Delta \tilde R_{NS}(x) \right) \otimes \Delta
\tilde q_{NS}(x,t)\;,  \label{e:NS}
\end{equation}
while for the Singlet and Gluon distributions we have:
\begin{equation}
\frac{\partial}{\partial t} \left (
\begin{array}{c}
\Delta \tilde q_{S}(x,t) \\
\Delta \tilde g(x,t)
\end{array}
\right) = \left [ \left(
\begin{array}{cc}
\Delta \tilde P^{(0)}_{qq}(x) & \Delta \tilde P^{(0)}_{qg}(x) \\
\Delta \tilde P^{(0)}_{gq}(x) & \Delta \tilde P^{(0)}_{gg}(x)
\end{array}
\right) + \alpha(t) \left(
\begin{array}{cc}
\Delta \tilde R_{qq}(x) & \Delta \tilde R_{qg}(x) \\
\Delta \tilde R_{gq}(x) & \Delta \tilde R_{gg}(x)
\end{array}
\right) \right] \otimes \left (
\begin{array}{c}
\Delta \tilde q_{S}(x,t) \\
\Delta \tilde g(x,t)
\end{array}
\right) \,,  \label{e:SG}
\end{equation}
where
\begin{equation}
\Delta\tilde R_{ij}(x) \equiv \Delta \tilde P^{(1)}_{ij}(x) - \frac{\beta_1}{%
2\beta_0}\Delta \tilde P^{(0)}_{ij}(x)\;.  \label{e:deltaR}
\end{equation}
In these equations the symbol $\otimes$ stands for
\begin{equation}
f(x) \otimes g(x) \equiv \int_x^1 \frac{dy}{y} f\left(\frac{x}{y}\right)
g(y)\;,  \label{e:conv}
\end{equation}
and
\begin{equation}
\tilde f(x) \equiv x f(x) \;.
\end{equation}
Instead of $Q^2$, we have used the variable $t$ defined by
\begin{equation}
t = -\frac{2}{\beta_0}ln\left[\frac{\alpha_s(Q^2)}{\alpha_s(Q_0^2)}\right]\;,
\end{equation}
where $\alpha_s$ is strong running coupling constant corrected at NLO:
\begin{equation}
\alpha_s(Q^2)= \frac{4 \pi}{\beta_0\,ln(Q^2/\Lambda^2)} \left[ 1- \frac{%
\beta_1\,ln(ln(Q^2/\Lambda^2))}{\beta_0^2\,ln(Q^2/\Lambda^2)} \right]\;,
\end{equation}
and so in Eqs.~(\ref{e:NS})-(\ref{e:SG})
\begin{equation}
\alpha(t) \equiv \frac{\alpha_s(Q^2_0)}{2\pi}Exp\left\{-\frac{\beta_0}{2}%
t\right\}\;.
\end{equation}
The explicit expressions for $\beta_0$, $\beta_1$ as well as for the
Splitting Functions $\Delta P_{ij}(x)$ can be found in \cite
{SplittingFunctions}.

The equations (\ref{e:NS}) and (\ref{e:SG}) can be written in the following
general form:
\begin{equation}
\frac{\partial}{\partial t} {\bf f}(t) = {\bf \Omega}(t) \odot {\bf f}(t)
\label{e:GF}
\end{equation}
where ${\bf f}(t)$ indicates the ``vector of components $f(x,t)$" and ${\bf %
\Omega}(t)$ a linear operator acting as:
\begin{equation}
\left [ {\bf \Omega}(t)^{~}_{~} \odot {\bf f}(t)_{}\right ]_x \equiv
\int_x^1 ~dy~ \omega(x,y,t)\;f(y,t)\;.  \label{e:defodot}
\end{equation}
Note that, in the Singlet-Gluon case, ${\bf f}(t)$ becomes a doublet of
vectors and ${\bf \Omega}(t)$ a $2~$x$~2$ matrix of operators.

Due to the logarithmic dependence of $t$ on $Q^2$, the range of values of
physical interest for $t-t_0$ ($t_0$ is the starting values of $t$, where
the parton distributions are assumed known) is small enough to expect that
the Taylor's series of the solution ${\bf f}(t)$ converges rapidly. On the
other hand, by deriving repeatedly the Eq.~(\ref{e:GF}) we can write:
\begin{equation}
\left . \frac{\partial^{k}}{\partial t^k} {\bf f}(t) \right |_{t=t_0}= ~{\bf %
M}^{(k)} \odot {\bf f}(t_0)
\end{equation}
where the operators ${\bf M}^{(k)}$ can be obtained recursively:
\begin{eqnarray}
{\bf M}^{(0)} & = & {\bf I}  \nonumber \\
{\bf M}^{(1)} & = & {\bf \Omega}_0  \nonumber \\
{\bf M}^{(2)} & = & {\bf \Omega}_0^{(1)} + {\bf \Omega}_0 \odot {\bf M}^{(1)}
\nonumber \\
{\bf M}^{(3)} & = & {\bf \Omega}_0^{(2)} + 2~{\bf \Omega}_0^{(1)} \odot {\bf %
M}^{(1)} + {\bf \Omega}_0 \odot {\bf M}^{(2)}  \nonumber \\
... & & .......  \nonumber \\
{\bf M}^{(k)} & = & \sum_{i=0}^{k-1}c_i^{(k)} {\bf \Omega}_0^{(k-1-i)} \odot
{\bf M}^{(i)}\,.  \label{e:iter}
\end{eqnarray}
The $c_i^{(k)}$ indicates the $i-$th term of the $k-$th row of Tartaglia
triangle and
\begin{equation}
{\bf \Omega}_0 \equiv {\bf \Omega}(t_0)\;, ~~~~~~~~~~~~~~~~~~~ {\bf \Omega}%
_0^{(k)} \equiv \left . \frac{\partial^{k}}{\partial t^k} {\bf \Omega}(t)
\right |_{t=t_0}\;.
\end{equation}
Then the solution can be written as:
\begin{equation}
{\bf f}(t) = \left ( \sum_{k=0}^{\infty} \frac{(t-t_0)^k}{ k !} {\bf M}%
^{(k)}\right ) \odot {\bf f}(t_0) \equiv {\bf T}(t-t_0) \odot {\bf f}(t_0)\;,
\label{e:solution}
\end{equation}
with ${\bf T}(t-t_0)$ the Evolution Operator.

As we will point out in section {\bf \ref{s:xintegration}} the
series in Eq.~(\ref{e:solution}) converge quickly enough to obtain
a very good approximation retaining only a first few terms. It is
worth to note that if the operator ${\bf \Omega}(t)$ can be
written as $h(t)\;{\bf \Omega^{\prime}} $ (with $h(t)$ a numerical
function) it is easy to show that the series in
Eq.~(\ref{e:solution}) reduces to:
\begin{equation}
{\bf f}(t) = Exp\left\{\left[\int_{t_0}^t h(\tau) d\tau \right ] {\bf %
\Omega^{\prime}} \right\} \odot {\bf f}(t_0)\;.  \label{e:esatta}
\end{equation}
This is the case of DGLAP equation at Leading Order (LO) approximation.
Nevertheless, in Eqs.~(\ref{e:NS})-(\ref{e:SG}), where NLO corrections are
included, we have ${\bf \Omega}(t) = {\bf \Omega}_1 + \alpha(t) {\bf \Omega}%
_2\;,$ with ${\bf \Omega}_1$ and ${\bf \Omega}_2$ non-commuting operators.
As a consequence the series in Eq.~(\ref{e:solution}) cannot be summed and
it is not possible write the solution in a closed form.

In the next section we show how to represent the operators ${\bf M}^{(k)}$
in a form suitable for numerical calculations.

\section{The $x$-Integration}

\label{s:xintegration}

The integrals in Eq.~(\ref{e:conv}) are evaluated with a method that
generalizes the one proposed in Ref. \cite{Fasching}. The method consists to
treat {\it exactly} the ``bad" behaviour of the kernel $\omega(x,y,t)$ in
Eq.~(\ref{e:defodot}) and {\it approximate} the ``smooth" function $f(y,t)$.
In particular, we construct a $M+1$ points grid ($x_0>0,x_1,...,x_{M-1},x_M=1
$) in the interval $]0,1]$ and approximate $f(x)$ (here we omit for brevity
the $t$ dependence) in each subinterval as linear combination of a suitable
set of function $g_l(x)$ (we have used the power series $x^{l-1}$):
\begin{equation}
f(x) \approx \sum_{l=1}^{m} a_l^{(k)} g_l(x)~~~~~~~~~~~~~~~ \forall x \in
\left[x_k,x_{k+1}\right]\;.  \label{e:fapprox}
\end{equation}
Here $a_l^{(k)}$ are coefficients which depend linearly on the values $f(x_k)
$ of $f(x)$ computed in the $x_k$ points; they can be obtained solving $%
\forall k$ the linear system of $m$ equations,
\begin{equation}
f(x_{k+r-m/2}) =
\sum_{l=1}^{m}a_l^{(k)}g_l(x_{k+r-m/2})~~~~~~~~~~~~~~~r\in\{1,...,m\}
\end{equation}
giving
\begin{equation}
a_l^{(k)} = \sum_{r=1}^{m} G^{(k)}_{lr} f(x_{k+r-m/2})\; .  \label{e:alk}
\end{equation}
In our numerical calculations we choose $m=4$ which corresponds to
approximate $f(x)$ in each interval $[x_k,x_{k+1}]$ by the cubic
which fits the four point $f(x_i)$, with $i=k-1,k,k+1,k+2$. In the
subintervals near the end-points ($x_0$, and $1$) we have reduced
the order $m$.

The general structure of the Polarized Splitting Functions which appear in
Eqs.~(\ref{e:NS})-(\ref{e:SG}) is\footnote{%
The same structure, however, holds for unpolarized and transversely
polarized splitting functions.}:
\begin{equation}
\Delta\tilde P(x) = \frac{{\cal A}(x)}{(1-x)_+} + {\cal B}(x) + \delta(1-x)%
{\cal C}\;,  \label{e:dp}
\end{equation}
where
\begin{equation}
\int_0^1 \frac{dz}{(1-z)_+}f(z) \equiv \int_0^1 dz\frac{f(z)-f(1)}{1-z}\,,
\end{equation}
and therefore the ``$i$ component" of the convolution is:
\begin{eqnarray}
\Delta\tilde P(x_i) \otimes f(x_i) & = & x_i\left(\int_{x_i}^1 \frac{dy}{y}
\frac{{\cal A}(x_i/y)f(y)- {\cal A}(1)f(x_i) }{ y - x_i } + \int_{x_i}^1
\frac{dy}{y^2} {\cal B}\left(\frac{x_i}{y}\right)f(y) \right) +  \nonumber \\
& & \left( {\cal C} + {\cal A}(1)ln(1-x_i) \right ) f(x_i)\;.
\label{e:dp-App}
\end{eqnarray}
Substituting Eq.~(\ref{e:fapprox}) in Eq.~(\ref{e:dp-App}) we obtain $%
\forall i \in \{0,...,~M-1\}$ ($~\sum_{k=M}^{M-1}\equiv 0$ is understood)
\begin{eqnarray}
\Delta\tilde P(x_i) \otimes f(x_i) & = & \sum_{l=1}^{m}
a_l^{(i)}\left(\beta_l^{i} + \rho_{il}^{i}\right) + \sum_{k=i+1}^{M-1}\;\;
\sum_{l=1}^{m} a_l^{(k)} \left(\gamma_{kl}^{i} + \rho_{kl}^{i} \right)
\nonumber \\
& + & \left( {\cal C} + {\cal A}(1)ln(1-x_i) - {\cal A}(1) \sigma^{i} \right
) f(x_i)\;;  \label{e:dpifi}
\end{eqnarray}
then Eq.~(\ref{e:alk}) implies:
\begin{equation}
\Delta\tilde P(x_i) \otimes f(x_i) = \sum_{k=0}^{M}\omega_{ik}f(x_k)
\label{e:final}
\end{equation}
where $\omega$ is the matrix of the coefficients of $f(x_k)$. The matrices $%
\beta$, $\gamma$, $\rho$ and $\sigma$ are given in the Appendix.

Due to the $ln(1-x_i)$ the Eq.~(\ref{e:dp-App}) contains a
logarithmic divergence for $i=M$. This is not a problem if the
solution $f(x,t)$ decreases to $0$ in $x=1$ faster than the
logarithm: in this case it's equivalent to put
$\omega_{Mk}=0~~~\forall k$ in Eq.~(\ref{e:final}), but in general
a numerical regularizator should be used to control the
$\lim_{x\to 1}ln(1-x)f(x,t)$.

Therefore the Eqs.~(\ref{e:NS})-(\ref{e:SG}) became \footnote{%
Note that $\omega^{(1)}$ matrices correspond to the convolutions of the
parton distributions with $\Delta \tilde R$ (cf Eq.~(\ref{e:deltaR})).}:
\begin{equation}
\frac{\partial}{\partial t} \Delta \tilde q_{NS}(x_i,t) = \sum_{k=0}^M\left(
\omega^{(0)~NS}_{ik} + \alpha(t)\omega^{(1)~NS}_{ik} \right) \Delta \tilde
q_{NS}(x_k,t)  \label{e:NS-om}
\end{equation}
\begin{equation}
\frac{\partial}{\partial t} \left (
\begin{array}{c}
\Delta \tilde q_{S}(x_i,t) \\
\Delta \tilde g(x_i,t)
\end{array}
\right) = \sum_{k=0}^{M}\left [ \left(
\begin{array}{cc}
\omega^{(0)~qq}_{ik} & \omega^{(0)~qg}_{ik} \\
\omega^{(0)~gq}_{ik} & \omega^{(0)~gg}_{ik}
\end{array}
\right) + \alpha(t) \left(
\begin{array}{cc}
\omega^{(1)~qq}_{ik} & \omega^{(1)~qg}_{ik} \\
\omega^{(1)~gq}_{ik} & \omega^{(1)~gg}_{ik}
\end{array}
\right) \right] \left (
\begin{array}{c}
\Delta \tilde q_{S}(x_k,t) \\
\Delta \tilde g(x_k,t)
\end{array}
\right)\;.  \label{e:SG-om}
\end{equation}
We solve these equations by means of the method shown in section {\bf \ref
{s:tevolution}}: the operator $\Omega(t)$ and then the ${\bf M}^{(k)}$
became now numerical matrices, and the symbol $\odot$ stands for the usual
rows by columns product. We would stress the fact that the matrices ${\bf M}%
^{(k)}$ depend only on the points $x_i$ and so they can be numerically
evaluated once for all.

\section{Numerical Analysis}

\label{s:nunan}

The convergence of our algorithm is controlled by two parameters: the order $%
n$ of the truncated series
\begin{equation}
{\bf T}^{(n)}(t-t_0) = \sum_{k=0}^{n} \frac{(t-t_0)^k}{ k !} {\bf M}^{(k)}\;,
\end{equation}
which define the evolution operator, and the number $M$ of the points of $x-$%
integration.

\begin{figure}[t]
\epsfig{file=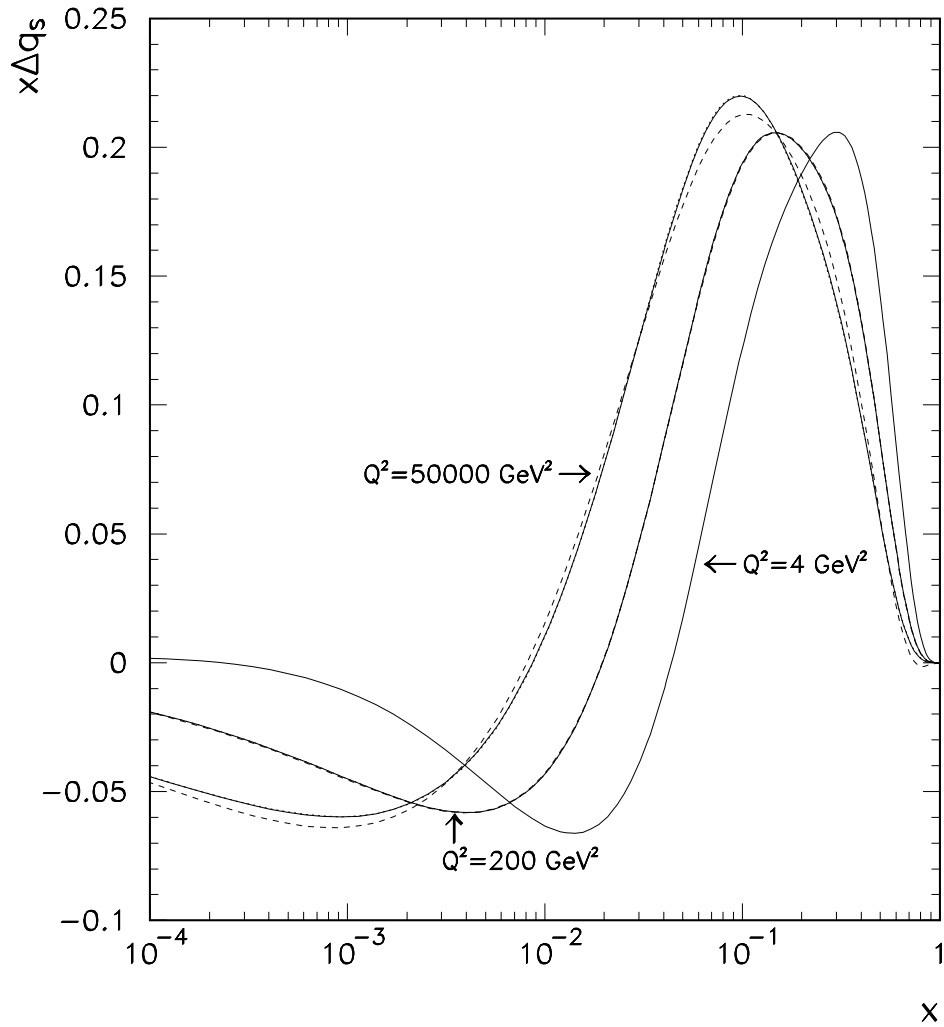,height=15cm}
\caption{ The initial Singlet distribution ($Q^2=4~GeV^2$, solid line) and
the evolved ones for $n=3$ (dashed lines), $n=6$ (dotted lines) and $n=12$
(solid lines) corresponding at $Q^2=200~GeV^2$ and $Q^2=50000~GeV^2$. We use
$M=100$. }
\label{f:fig1}
\end{figure}

To test the accuracy of our method we evolve the Gehrmann and Stirling
polarized singlet-gluon initial distributions (cf \cite{GS}):
\begin{eqnarray}
\Delta \tilde q_{S}(x,t_0) & = & x\Delta u_v + x\Delta d_v + 6x \Delta S
\nonumber \\
& = & .918\ast 1.365 \ast x^{0.512}(1-x)^{3.96} (1+11.65 x-4.6 \sqrt{x})
\nonumber \\
& - & .339\ast 3.849 \ast x^{.78}(1-x)^{4.96} (1+7.81 x-3.48 \sqrt{x})
\nonumber \\
& - & 6\ast .06 \ast 18.521\ast x^{.724}(1-x)^{14.4}(1+4.63x -4.96 \sqrt{x})
\\
\Delta \tilde g(x,t_0) & = &1.71 \ast 3.099 \ast x^{.724}(1-x)^{5.71}
\end{eqnarray}
from $Q_0^2 = 4~GeV^2~(t_0=0)$ to $Q^2 =200~GeV^2~(t=0.136)$ and $%
Q^2=50000~GeV^2~(t=0.245)$. We choose to work, as in the paper \cite{Kumano}%
, in the fixed flavour scheme, $n_f=3$, with $\Lambda^{(4)}_{QCD}=231~MeV$,
and without taking into account, in the $Q^2$ evolution of $\alpha_s$, quark
thresholds. The range $]0,1]$ has been divided in $M$ steps by $M+1$ points:
$x_0, x_1, ..., x_M$ distributed in such a way that the function $ln(x) + 2x$
varies by the same amount at any step; this function is slightly different
from the pure logarithmic distribution commonly used in literature \cite
{Kumano,Fasching}, but allow, in our case, a more uniform distribution of
the numerical errors. The end points are fixed to be $x_0=1\times 10^{-8}$
and $x_M=1$; however, for a better reading, in the Figures~{\bf \ref{f:fig1}}%
$-${\bf \ref{f:fig4}} the $x-$axis ranges from $10^{-4}$ to $1$.

\begin{figure}[t]
\epsfig{file=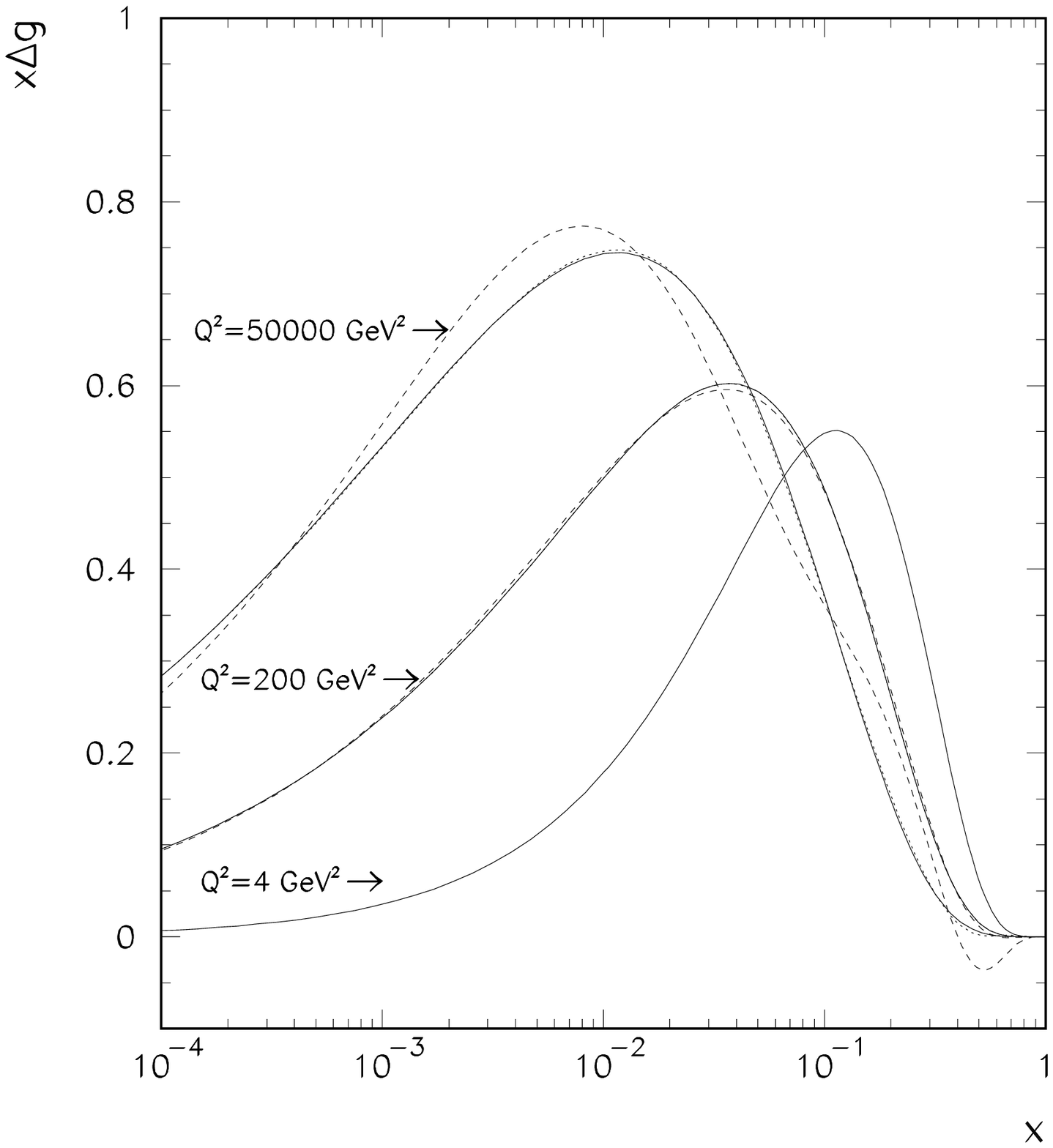,height=15cm}
\caption{The same in Fig. \ref{f:fig1} for Gluons.}
\label{f:fig2}
\end{figure}

First, we fix $M= 100$. In Figs. {\bf \ref{f:fig1}}$-${\bf \ref{f:fig2}} are
reported the evolved singlet and gluon distributions, respectively, obtained
with $n=3,~6~$and$~12$ for $Q^2 =200~GeV^2$ and $Q^2 =50000~GeV^2$. It is
worth to note the very fast convergence of the series to the solution, as
already observed above. As a matter of fact, the maximum difference between
the solutions relative to $n=6$ and $n=12$ is $1.4\times 10^{-5}~$($%
7.6\times 10^{-4}$) for the singlet, and $9.3\times 10^{-5}~$($5.3\times
10^{-3}$) for the gluon distribution, in correspondence of $Q^2=200~GeV^2~$($%
Q^2=50000~GeV^2$).

\begin{figure}[t]
\epsfig{file=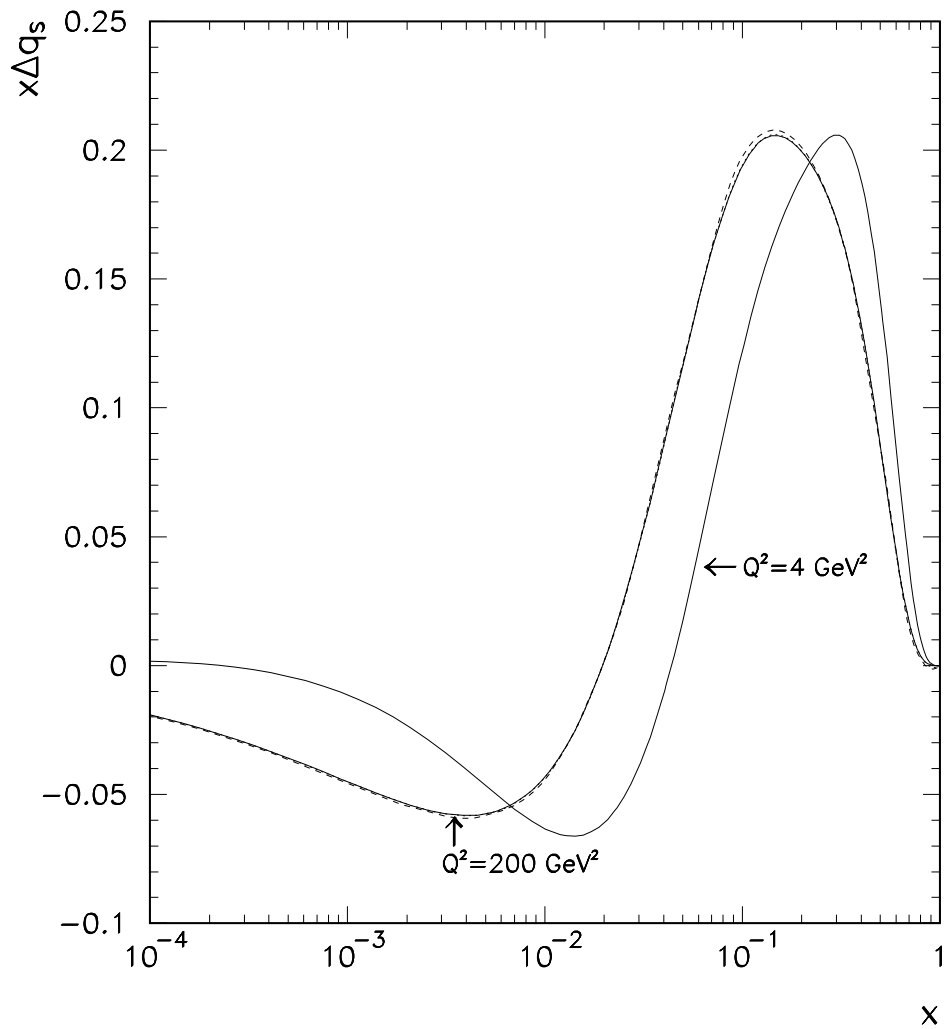,height=15cm}
\caption{ The initial Singlet distribution ($Q^2=4~GeV^2$, solid line) and
the evolved one at $Q^2=200~GeV^2$ with $M=100$ (solid line), $M=50$ (dotted
line) and $M=25$ (dashed line) with $n=12$. }
\label{f:fig3}
\end{figure}

\begin{figure}[t]
\epsfig{file=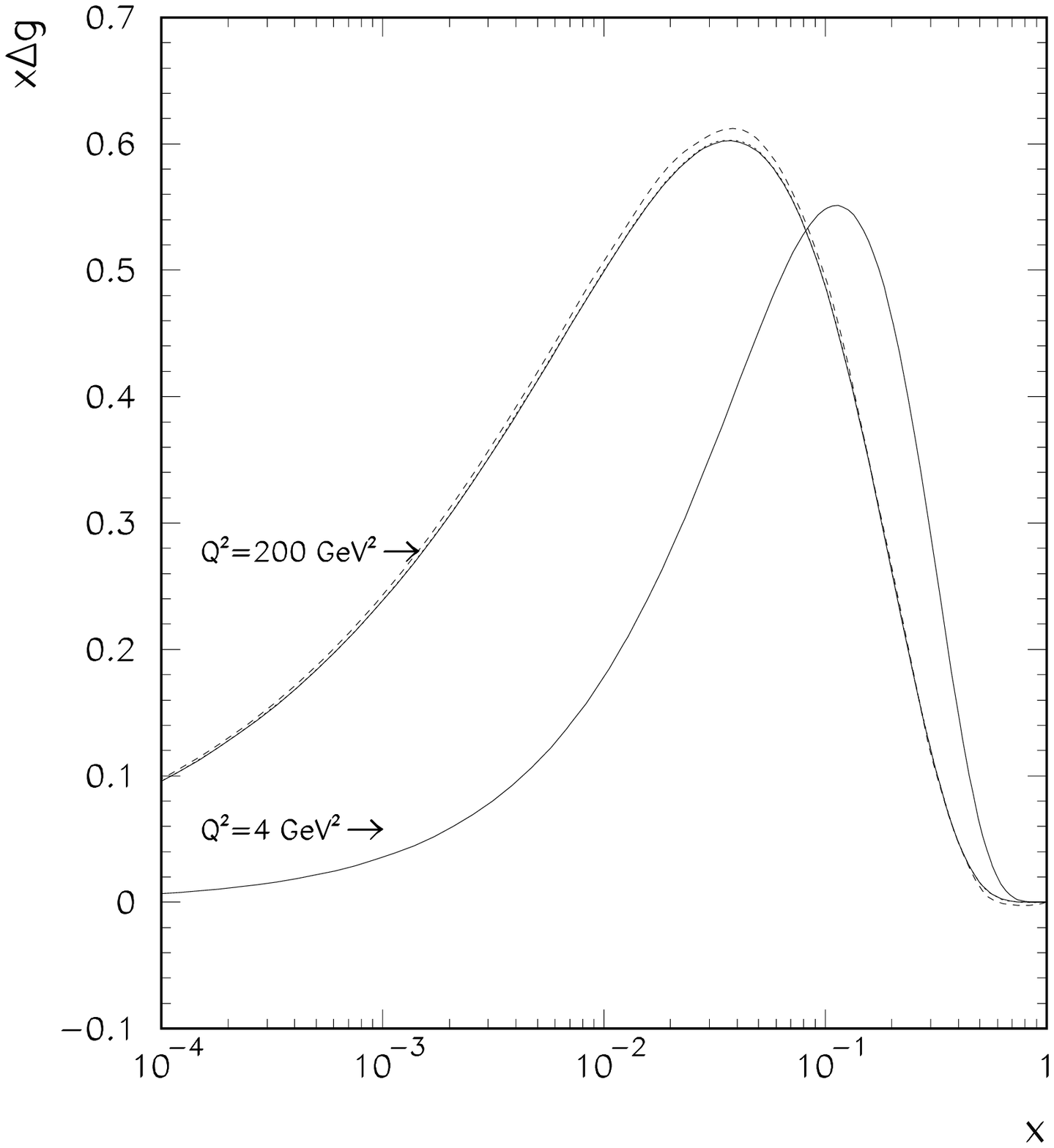,height=15cm}
\caption{ The same in Fig. \ref{f:fig3} for Gluons. }
\label{f:fig4}
\end{figure}

Next we fix $n=12$ and $Q^2= 200~GeV^2$. In Figs. {\bf \ref{f:fig3}}$-${\bf
\ref{f:fig4}} are plotted the approximated evolved distributions with $%
M=25,~50,~100$: the maximum difference on the common points between $M=50$
and $M=100$ is $4.6 \times 10^{-4}$ for the singlet and $7.9 \times 10^{-4}$
for the gluons. By comparing the results in Fig. {\bf \ref{f:fig3}}$-${\bf
\ref{f:fig4}} with the corresponding Figs. {\bf 1}-{\bf 4} in Ref. \cite
{Kumano}, we observe, besides a good numerical agreement of the results, a
faster convergence as the number M of integration points increases, as a
consequence of our more accurate $x$-integration procedure with respect to
the so called ``brute force'' methods. In fact it should be observed that
reducing from $m=4$ (cubic) to $m=2$ (linear) the order of approximation of $%
f(x)$ in Eq.~(\ref{e:fapprox}), the accuracy ${\cal E}(x,t)$ (defined in the
sequel) becomes about 1 and 3 order of magnitude bigger, respectively for
singlet and gluon.

\begin{figure}[t]
\epsfig{file=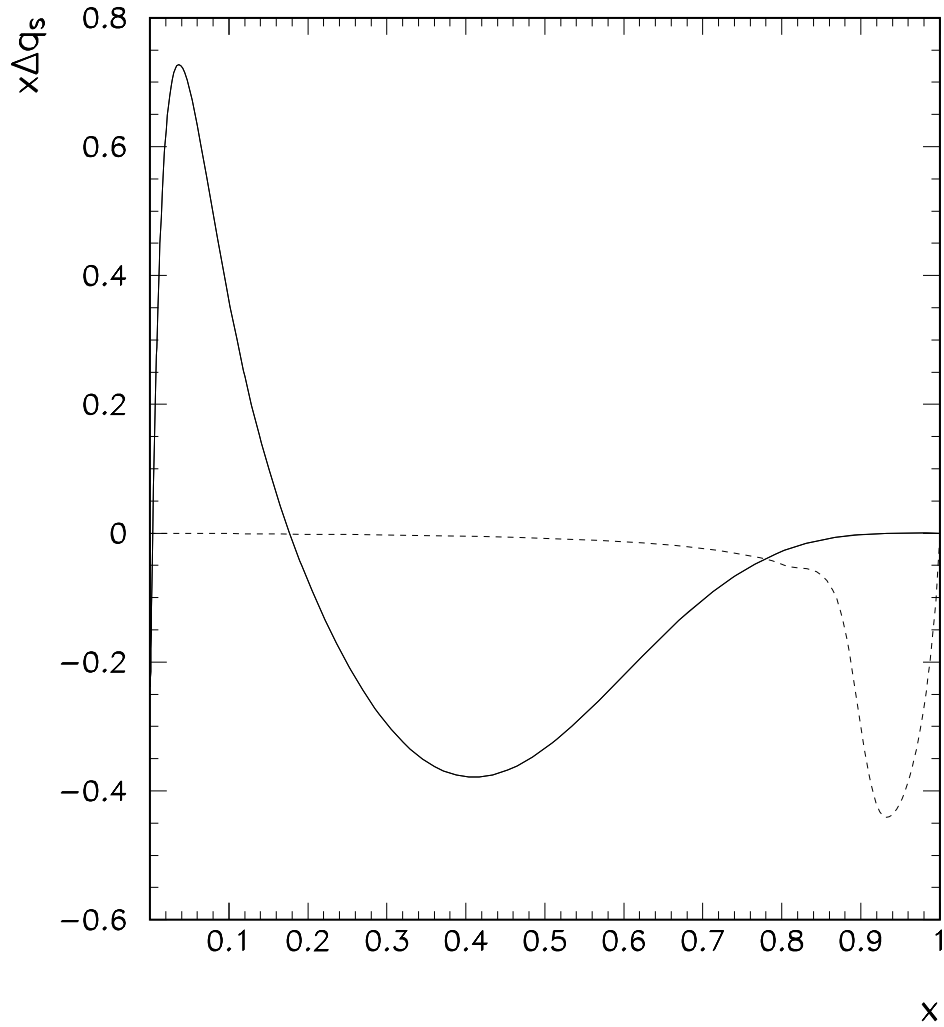,height=15cm}
\caption{ For $n=12$, $M=100$ and $Q^2 = 200~GeV^2$ both sides of the Eq.~(%
\ref{e:SG}) are plotted (solid line and dotted line), in correspondence of
the Singlet distribution. Dashed line represents ${\cal E}(x)~\times~10^{3}$
(see text).}
\label{f:fig5}
\end{figure}

\begin{figure}[t]
\epsfig{file=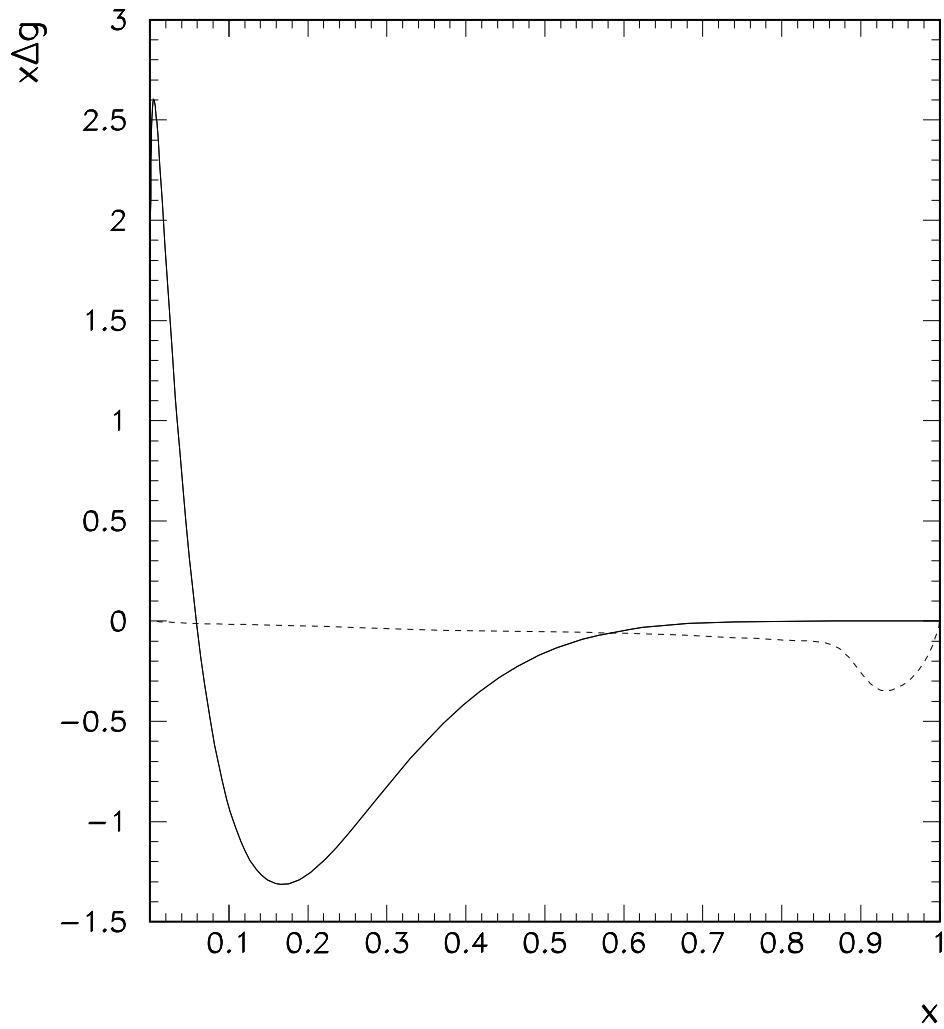,height=15cm}
\caption{The same in Fig. \ref{f:fig5} for Gluons. Dashed line represents $%
{\cal E}(x)~\times~10^{4}$ (see text).}
\label{f:fig6}
\end{figure}

%%%%%%%%%%%%%%%%%%%%%%%%%%%%%%%%%%%%%%%%%%%%%%%%%%%%%%%%%%%%%%%%%%%%%
%To complete the error analysis, let us introduce a global accuracy
%%%%%%%%%%%%%%%%%%%%%%%%%%%%%%%%%%%%%%%%%%%%%%%%%%%%%%%%%%%%%%%%%%%%%
To discuss the degree of accuracy of our method, let us introduce a global
accuracy ${\cal E}(x,t)$ defined as the difference between left and
right-hand side of the Eq.~(\ref{e:SG}). The comparison between the range of
values of ${\cal E}$ with the one of both sides of Eq.~(\ref{e:SG})
represents a very good estimate of the degree of accuracy of the solution.
In Figs. {\bf \ref{f:fig5}} and {\bf \ref{f:fig6}} are plotted, for $n=12$, $%
M=100$ and $Q^2 = 200~GeV^2$ both sides of the Eq.~(\ref{e:SG}) and the
corresponding (rescaled) accuracy ${\cal E}(x,t)$ \footnote{%
Note that the integration in the right-hand side has been performed
numerically after an $x$-interpolation of the discrete values obtained with
the evolution operator, while the left-hand side is worked out by direct
derivation of Eq.~(\ref{e:solution}).}. It appears evident that an excellent
approximation of the solution is obtained.

Another advantage of our method, once fixed the accuracy of the solution,
appears to be the running time to get each evolution. In fact, the simple
analytical structure of the evolution matrix ${\bf T}$ makes the solution
procedure considerably fast. As a matter of fact, once given the Splitting
Functions and constructed the corresponding matrices ${\bf M}^{(k)}$ (we
have used Mathematica \cite{Mathematica} to do this), a single evolution,
i.e. the multiplication of the ${\bf T}$ evolution matrix by the initial
vector, require, for $n=12$ and $M=100$, about $6 \times 10^{-2}~sec$ on an
AlphaServer 1000 using a Fortran Code.

Particularly interesting is the comparison between our method and
the one presented in \cite{FurmPetr,CorianoSavkli}, where an
evolution operator is also introduced. Firstly we observe that the
latter method is based on a polynomial expansion of the splitting
and distribution functions. The expansion is equivalent to an
expansion in power of $x$. As a consequence it is affected by
problems of convergence for $x\rightarrow 0$, due to the branch
point in zero of the involved functions. This is the source of the
difficult encountered in the small-$x$ region, which are not
present in our approach in which an optimized Newton-Cotes-like
quadrature formula is employed.\newline Second, also in the
$x$-region of convergence, the Laguerre polynomial expansion
needs, for each evolution process, the computation of the moments
of the initial parton distributions with respect to the
polynomials: this procedure requires a remarkable amount of
CPU-time with respect to our approach in which only the evaluation
of the initial parton distribution in the $M$ grid points is
needed.

\section{Conclusions}

\label{s:conclu}

We have proposed a new algorithm to solve the DGLAP evolution equations, in
the $x$-space, which appears suitable for a rather large class of coupled
integrodifferential equations.

The method produces a solution which is analytical in the
$Q^2$-evolution parameter and approximate, but rapidly convergent,
in the $x-$space. It allows to construct, once for all, an
evolution operator in matrix form. It depends only on the
splitting functions appearing in the equations and can be rapidly
applied to whatever initial distribution to furnish the evolved
one, requiring for each evolution only a few hundredth of second.

It is worth to note the reliability of our $x-$integration algorithm, which
gets excellent approximations on the whole $x-$range (we use, for all the
calculations, $10^{-8}\le$ $x$ $\le$ 1), also with few integration points,
resulting in an evolution matrix of particularly small dimensions.

In conclusion, our method, whose numerical implementation is
straightforward, appears to be very fast, very accurate and extremely stable
with respect to the increasing of convergence parameters (i.e. $n$, the
order of the truncated series which gives the Evolution Operator, and $M$,
the number of integration points). For these reasons it represents a
powerful tool to analyze the experimental data on nucleon structure
functions.

\vspace{1truecm}

{\large {\bf Acknowledgments} } \newline

Many thanks to Mario Pellicoro for helpful discussions on
numerical problems.

%\vspace{.5truecm}

\newpage

 \setcounter{equation}{0}
 \renewcommand{\theequation}{A.\arabic{equation}} %
 \addtocounter{appendix}{1}
 \vspace{.7truecm}
 \noindent {\Large {\bf Appendix}}
 \vspace{.5truecm}

\noindent {} Here we give the definitions $\beta$, $\gamma$,
$\rho$ and $\sigma$ matrices appearing in Eq.~(\ref{e:dpifi})
\begin{eqnarray}
\beta^{(i)}_l & \equiv & x_i \int_{x_i}^{x_{i+1}} \frac{dy}{y} \frac{{\cal A}%
(x_i/y)g_l(y)- {\cal A}(1)g_l(x_i)}{y-x_i} \\
\rho^{(i)}_{kl} & \equiv & x_i \int_{x_k}^{x_{k+1}} \frac{dy}{y^2} {\cal B}%
(x_i/y)g_l(y) \\
\gamma^{(i)}_{kl} & \equiv & x_i \int_{x_k}^{x_{k+1}} \frac{dy}{y} \frac{%
{\cal A}(x_i/y)g_l(y)}{y-x_i} \\
\sigma^{(i)} & \equiv & x_i \int_{x_{i+1}}^{1} \frac{dy}{y(y-x_i)}
\end{eqnarray}
where ${\cal A}$, ${\cal B}$ and ${\cal C}$ are the coefficients
appearing in the Eq.~(\ref{e:dp}). Note that, choosing $g_l(x) =
x^{l-1}$ it is possible to analytically evaluate all the previous
integrals in term of logarithms and $Li_n(z)=
\sum_{j=1}^{\infty}z^j/j^n$; the final expressions are too long
and are not reported here.

\vspace{0.5truecm}

%\newpage

\end{document}